\documentclass{siamltex}\usepackage{url}  

\usepackage{amsmath,amssymb,amsfonts}   
\usepackage{graphicx,bm}
\usepackage[colorlinks=true]{hyperref}
 \hypersetup{urlcolor=blue, citecolor=red}
 \usepackage[latin1]{inputenc}

\newcommand{\calU}{{\mathcal U}}
\newcommand{\R}{{\mathbb R}}

\newcommand{\X}{\mathbf{X}}
\renewcommand{\P}{\mathbb{P}}
\newcommand{\x}{\mathbf{x}}
\newcommand{\y}{\mathbf{y}}
\newcommand{\e}{{\mathrm e}}
\newcommand{\E}{{\mathbb E}}

\newcommand{\n}{\mathbf n}
\newcommand{\calT}{{\mathcal T}}
\renewcommand{\P}{\mathbb P}
\newcommand{\p}{\widetilde{p}}

\begin{document}

\title{\bf  Asymptotic analysis of target fluxes in the three-dimensional narrow capture problem}


\author{Paul C. Bressloff\thanks{Department of Mathematics, University of Utah, Salt Lake City, UT 84112
USA ({\tt bressloff@math.utah.edu})}  }

 \maketitle


\begin{abstract} 
We develop an asymptotic analysis of target fluxes in the three-dimensional (3D) narrow capture problem. The latter concerns a diffusive search process in which the targets are much smaller than the size of the search domain. The small target assumption allows us to use matched asymptotic expansions and Green's functions to solve the diffusion equation in Laplace space. In particular, we derive an asymptotic expansion of the Laplace transformed flux into each target in powers of the non-dimensionalized target size $\epsilon$. One major advantage of working directly with fluxes is that one can generate statistical quantities such as splitting probabilities and conditional first passage time moments without having to solve a separate boundary value problem in each case. However, in order to derive asymptotic expansions of these quantities, it is necessary to eliminate Green's function singularities that arise in the limit $s\rightarrow 0$, where $s$ is the Laplace variable. We achieve this by considering a triple expansion in $\epsilon$, $s$ and $\Lambda\sim \epsilon /s$. This allows us to perform partial summations over infinite power series in $\Lambda$, which leads to multiplicative factors of the form $\Lambda^n/(1+\Lambda)^n $. Since $\Lambda^n/(1+\Lambda)^n \rightarrow 1$ as $s\rightarrow 0$, the singularities in $s$ are eliminated. We then show how corresponding asymptotic expansions of the splitting probabilities and conditional MFPTs can be derived in the small-$s$ limit. The resulting expressions agree with previous asymptotic expansions derived by solving separate boundary values problems for each statistical quantity, although the latter were only carried out to second order in the expansions. Here we also determine the third order contributions, which are $O(\epsilon^2)$ and $O(\epsilon)$ in the case of the splitting probabilities and conditional MFPTs, respectively. Finally, we illustrate the theory by considering a pair of targets in a spherical search domain, for which the Green's functions can be calculated explicitly. \end{abstract}


\section{Introduction}

 The classical narrow capture (or escape) problem concerns diffusive search processes where the targets are much smaller than the size of the search domain. This then allows matched asymptotic expansions and Green's functions to be used to solve the boundary value problems (BVPs) for the splitting probabilities and moments of the conditional FPT density \cite{Ward93,Schuss07,Bressloff08,Coombs09,Cheviakov11,Chevalier11,Holcman14,Coombs15,Ward15,Bressloff15,Lindsay15,Lindsay16,Lindsay17}. However, there are a growing number of search problems that require knowledge of the probability flux into each target, whose Laplace transform acts as a generator of the conditional FPT moments \cite{Bressloff20A}. For example, one way to increase the efficacy of a search process is to include a stochastic resetting protocol, whereby the position of the searcher is reset to a fixed location $\x_r$ at a random sequence of times, which is typically (but not necessarily) generated by a Poisson process. In many cases there exists an optimal resetting rate for minimizing the mean first passage time (MFPT) to reach a target, see the review \cite{Evans20} and references therein. If the particle loses all memory of previous search phases following resetting, then renewal theory can then be used to express statistical quantities with resetting in terms of the target fluxes without resetting \cite{Bressloff20A,Bressloff20B,Bressloff20m}. 
 
We recently used asymptotic methods to analyze target fluxes in the two dimensional (2D) narrow capture problem \cite{Bressloff20B}. We proceeded by Laplace transforming the forward diffusion equation, which was then solved by constructing an inner or local solution valid in an $O(\epsilon)$ neighborhood of each target, and then matching to an outer or global solution that is valid away from each neighborhood. The small dimensionless parameter $\epsilon$ represents the size of each target relative to the size of the search domain. The matching procedure leads to terms involving the 2D Green's function $G(\x,s|\x_0)$ of the modified Helmholtz equation $D\nabla^2 G-sG=-\delta(\x-\x_0)$, where $s$ is the Laplace variable. Since $G$ has a logarithmic singularity, $G(\x,s|\x_0)\sim -\ln|\x-\x_0|)$ as $\x\rightarrow \x_0$, one obtains an asymptotic expansion in $\nu= -1/\ln \epsilon$ rather than $\epsilon$ itself. Moreover, it is possible to sum over the logarithmic terms non-perturbatively, which is equivalent to calculating the asymptotic solution for all terms of $O(\nu^k)$ for any $k$  \cite{Ward93}. Having solved the diffusion equation in Laplace space, we determined the corresponding Laplace transform of the probability flux into each target. We then showed how taking the small-$s$ limit of the fluxes for fixed $\nu$ generated corresponding asymptotic expansions of the splitting probabilities and conditional FPT moments, without having to solve a separate boundary value problem for each one. In addition, for finite $s$ we used the flux expansions to determine the effects of stochastic resetting on the mean first passage time (MFPT) to find a target. (The Laplace variable is now identified with the resetting rate $r$.) In particular, we exploited the exponential-like asymptotic decay of the Green's function for the modified Helmholtz equation (see Ref. \cite{Lindsay16}), in order to construct boundary-free approximations of statistical quantities in the presence of resetting. This allowed us to identify target configurations where the MFPT is minimized at an optimal resetting rate.

In this paper, we develop a corresponding asymptotic analysis of target fluxes in the three-dimensional (3D) narrow capture problem. The main difference from the 2D case is that the Green's function singularity is now of the form $G(\x,s|\x_0)\sim 1/|\x-\x_0|$ as $\x\rightarrow \x_0$. This significantly alters the details of the analysis and leads to an asymptotic expansion of the Laplace transformed fluxes in powers of $\epsilon$. One major complicating factor in using these expansions to generate the conditional FPT moments is that one has to deal with the fact that the Green's function is singular in the limit $s\rightarrow 0$, that is, $G(\x,s|\x_0)\sim 1/s$. (This was not an issue in 2D because we could sum over all logarithmic terms and cancel the singularities). A major result of this paper is to show how one can eliminate the singularities by treating the asymptotic expansion of the Laplace transformed flux into each target as a triple expansion in $\epsilon$, $s$ and $\Lambda\sim \epsilon /s$. This allows us to perform partial summations over infinite power series in $\Lambda$, which leads to multiplicative factors of the form $\Lambda^n/(1+\Lambda)^n $. Since $\Lambda^n/(1+\Lambda)^n \rightarrow 1$ as $s\rightarrow 0$, the singularities in $s$ are eliminated. We then show how corresponding asymptotic expansions of the splitting probabilities and conditional MFPTs can be derived in the small-$s$ limit. The resulting expressions agree with previous asymptotic expansions derived by solving separate boundary values problems for each statistical quantity, although the latter were only carried out to second order in the expansions. Here we also determine the third order contributions, which are $O(\epsilon^2)$ and $O(\epsilon)$ in the case of the splitting probabilities and conditional MFPTs, respectively.

The structure of the paper is as follows. In \S 2 we define the 3D narrow capture problem and show how splitting probabilities and conditional FPTs are related to the target fluxes in Laplace space. In \S 3 we carry out the asymptotic analysis in the case of spherically-shaped targets, systematically matching inner and outer solutions. We thus derive an explicit expression for the fluxes to $O(\epsilon^3)$. In \S 4 we show how singularities in the limit $s\rightarrow 0$ can be eliminated by performing partial summations, and then use this to derive corresponding asymptotic expansions of the splitting probabilities and conditional MFPTs. We illustrate the theory in \S 5 by considering a pair of targets in a spherical search domain, for which the Green's functions can be calculated explicitly. We show the breakdown of the finite asymptotic expansion for small-$s$ and indicate how the fluxes for finite $s$ can be used to determine the splitting probabilities in the presence of stochastic resetting. Finally, in \S 6 we briefly indicate how to extend the analysis to non-spherical targets.

\section{Narrow capture problem in 3D}

\begin{figure}[b!]
\begin{center} 
\includegraphics[width=7cm]{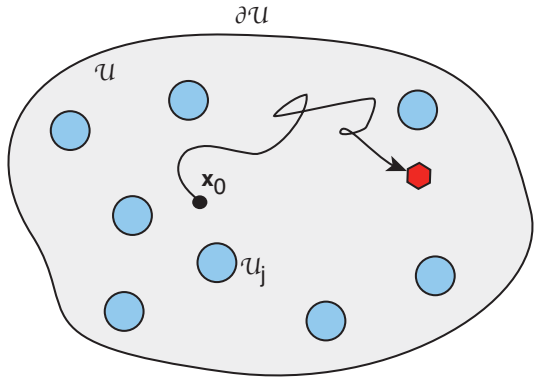} 
\caption{Diffusion of a particle in a 3D domain $\calU\subseteq \R^3$ with $N$ small targets $\calU_j$, $j=1,\ldots,N$. If the particle hits a boundary $\partial\calU_j$ then it is absorbed. The initial position is denoted by $\x_0$.}
\label{fig1}
\end{center}
\end{figure}

Consider a three-dimensional bounded domain $\calU\subset \R^3$ that contains a set of $N$ small interior targets $\calU_k$, $k=1,\ldots,N$, with $\bigcup_{j=1}^N \calU_k=\calU_a\subset \calU$, see Fig. \ref{fig1}. 
Let $p(\x,t|\x_0)$ be the probability density that at time $t$ a particle is at $\X(t)=\x$, having started at position $\x_0$. Then
\begin{subequations} 
\label{master}
\begin{align}
	\frac{\partial p(\x,t|\x_0)}{\partial t} &= D\nabla^2 p(\x,t|\x_0), \ \x\in \calU\backslash \calU_a,\quad \nabla p \cdot \n=0, \ \x\in \partial \calU,\\
	p(\x,t|\x_0) &=0,\  \x\in \partial\calU_a,
	\end{align}
	\end{subequations} 
together with the initial condition $p(\x,t|\x_0)=\delta(\x-\x_0)$.
Each target is assumed to have a size $|\calU_j|\sim \epsilon^3 |\calU|$ with $\calU_j\rightarrow \x_j\in \calU$ uniformly as $\epsilon \rightarrow 0$, $j=1,\ldots,N$. The targets are also taken to be well separated in the sense that $|\x_i-\x_j|=O(1)$, $j\neq i$, and $\mbox{dist}(x_j,\partial \calU)=O(1)$.
For the sake of illustration, we take each target to be a sphere of radius $\epsilon \ell_j$. Thus $\calU_i=\{\x \in \calU, \ |\x-\x_i|\leq \epsilon \ell_i\}$.

The probability flux into the $k$-th target at time $t$ is 
\begin{align}
\label{J}
	J_k(\x_0,t)&=-D \int_{\partial \calU_k} \nabla p(\x,t|\x_0)\cdot \n d\sigma,\ k = 1,\ldots,N,
	\end{align}
	where $\n$ is the inward normal to $\calU_k$.
Hence, the probability that the particle is captured by the $k$-th target after time $t$ is
\begin{equation}
\label{Pi}
\Pi_k(\x_0,t)=\int_t^{\infty}J_k(\x_0,t')dt' ,
\end{equation}
and the corresponding splitting probability is
\begin{equation}
\label{split}
\pi_k(\x_0)=\Pi_k(\x_0,0)=\int_0^{\infty}J_k(\x_0,t')dt' =\widetilde{J}_k(\x_0,0).
\end{equation}
It follows that the Laplace transform of $\Pi(\x_0,t)$ is given by
\begin{align}
\label{Pik}
s\widetilde{\Pi}_k(\x_0,s)-\pi_k=-\widetilde{J}_k(\x_0,s)=D\int_{\partial \calU_k} \nabla \widetilde{p}(\x,s|\x_0)\cdot \n d\sigma.
\end{align}
Next we introduce the survival probability that the particle hasn't been absorbed by a target in the time interval $[0,t]$, having started at $\x_0$:
\begin{equation}
\label{Q1}
Q(\x_0,t)=\int_{\calU\backslash \calU_a}p(\x,t|\x_0)d\x.
\end{equation}
Differentiating both sides of this equation with respect to $t$ and using equations (\ref{master}) implies that
\begin{align}
\frac{\partial Q(\x_0,t)}{\partial t}&=D\int_{\calU\backslash \calU_a}\nabla\cdot \nabla p(\x,t|\x_0)d\x=D\sum_{k=1}^N \int_{ \partial \calU_k}\nabla p(\x,t|\x_0)\cdot \n d\sigma\nonumber \\
& =-\sum_{k=1}^NJ_k(\x_0,t).
\label{Q2}
\end{align}
Laplace transforming equation (\ref{Q2}) gives
\begin{equation}
\label{QL}
s\widetilde{Q}(\x_0,s)-1=- \sum_{k= 1}^N \widetilde{J}_k(\x_0,s).
\end{equation}
We have used the initial condition $Q(\x_0,0)=1$. It immediately follows from equation (\ref{split}) that
\begin{equation}
\label{split0}
\sum_{k=1}^N\pi_k(\x_0)=1 -\lim_{s\rightarrow 0}s\widetilde{Q}(\x_0,s)=1.
\end{equation}
In other words, in the case of a bounded domain $\calU$ with a reflecting boundary $\partial \calU$, the searcher eventually finds a target with probability one.

Since the probability of the particle being captured by the $k$-th target is typically less than unity ($\pi_k<1$), it follows that the moments of the corresponding FPT density are infinite unless we condition on the given event. The MFPT $\calT_k$ to be captured by the $k$-th target is given by
\begin{equation}
\calT_k(\x_0)=\inf\{t>0; \X(t)\in \partial\calU_k|\X(0)=\x_0\},
\end{equation}
with $\calT_k=\infty$ if the particle is captured by another target. Introducing the set of events $\Omega_k=\{\calT_k<\infty\}$, we can then define the conditional FPT density according to
\begin{align*}
f_k(\x_0,t)dt&=\P[t<\calT_k<t+dt|\calT_k<\infty,\X(0)=\x_0]\\
&=\P[t<\calT_k<t+dt|\X(0)=\x_0]/\P[\Omega_k] \nonumber \\
&=\frac{\Pi_k(\x_0,t]-\Pi_k(\x_0,t+dt)}{\pi_k(\x_0)}
=-\frac{1}{\pi_k[\x_0]}\frac{\partial \Pi_k(\x_0,t]}{\partial t} ,
\end{align*}
since $\pi_k=\P[\Omega_k]$.
That is,
\begin{equation}
\label{fk}
f_k(\x_0,t)=\frac{J_k(\x_0,t)}{\pi_k(\x_0)}.
\end{equation} 
Hence, the Laplace transform of $f_k(\x_0,t)$ is the generator of the moments of the conditional FPT density:
\begin{align}
\label{fkLT}
\E[\e^{-s\calT_k}|1_{\Omega_k}]=\widetilde{f}_k(\x_0,s)=\frac{\widetilde{J}_k(\x_0,s)}{\widetilde{J}_k(\x_0,0)},
\end{align}
and
\begin{equation}
T_k^{(n)}=\E[\calT_k^n|1_{\Omega_k}]=\left . \left (-\frac{d}{ds}\right )^n\E[\e^{-s\calT_k}|1_{\Omega_k}]\right |_{s=0}=\left . \left (-\frac{d}{ds}\right )^n\widetilde{f}_k(\x_0,s)\right |_{s=0}.
\end{equation}
In particular, using equations (\ref{Pik}) and (\ref{fkLT}), the first and second moments $T_k=T_k^{(1)}$ and $T_k^{(2)}$ are
\begin{equation}
\label{mfpt}
\pi_k(\x_0)T_k(\x_0)=-\pi_k(\x_0)\left .\frac{d\widetilde{f}_k(\x_0,s)}{ds}\right |_{s=0}=\widetilde{\Pi}_k(\x_0,0),
\end{equation}
and
\begin{equation}
\label{fpt2}
\pi_k(\x_0)T_k^{(2)}(\x_0)=\pi_k(\x_0)\left .\frac{d^2\widetilde{f}_k(\x_0,s)}{ds^2}\right |_{s=0}=-2\left . \frac{d\widetilde{\Pi}_k(\x_0,0)}{ds}\right |_{s=0}.
\end{equation}

\section{Matched asymptotics}

It follows from the above analysis that one way to calculate the splitting probabilities and the Laplace transformed conditional FPT densities (\ref{fkLT}) is to solve equation (\ref{master}) in Laplace space. The latter takes the form
\begin{subequations} 
\label{masterLT}
\begin{align}
	D\nabla^2 \p(\x,s|\x_0) -s\p(\x,s|\x_0)&= -\delta(\x-\x_0) , \ \x\in \calU\backslash \calU_a,\\
	 \nabla \p \cdot \n=0, \ \x\in \partial \calU;\  \p(\x,s|\x_0)&= 0,\ \x\in \partial\calU_a.
	\end{align}
\end{subequations}
Equations (\ref{masterLT}) define a boundary value problem that can be analyzed along analogous lines to previous studies of diffusion in 3D domains with small targets \cite{Cheviakov11,Coombs15}. We proceed by matching appropriate
`inner' and `outer' asymptotic expansions in the
limit of small target size $\varepsilon\to 0$. In the outer region, which is outside an $O(\epsilon)$ neighborhood of each trap, $q(\x,s)$ is expanded as
\[\p(\x,s|\x_0)\sim \p_0(\x,s|\x_0)+\epsilon\p_1(\x,s|\x_0)+\epsilon^2 \p_2(\x,s|\x_0)+\ldots
\]
The leading order term is the solution without any holes and satisfies
\begin{align}
\label{asym0}
D\nabla^2 \p_0-s\p_0&=-\delta(\x-\x_0),\, \x\in \calU;\ \nabla \p_0\cdot \n=0,\, \x\in \partial \calU,
\end{align}
That is, $\p_0=G(\x,s|\x_0)$
where $G$ is the Neumann Green's function of the modified Helmholtz equation. In particular.
\begin{subequations}
\label{GMH}
\begin{align}
 \int_{\calU}G(\x,s|\x_0)d\x=\frac{1}{s};\
	 G(\x,s|\x_0)&=\frac{1}{4\pi D|\x-\x_0|}+R(\x,s|\x_0),
	\end{align}
	\end{subequations}
where $R$ is the regular part of $G$. In addition, for $n\geq 1$, we have
\begin{align}
\label{asym1}
D\nabla^2 \p_n-s\p_n&=0,\, \x\in \calU\backslash \{\x_1,\ldots,\x_N\};\ \nabla \p_n\cdot \n=0,\, \x\in \partial \calU,
\end{align}
together with certain singularity conditions as $\x\rightarrow \x_j$, $j=1,\ldots,N$. The latter are determined by matching to the inner solution.

\subsection{Leading-order contributions}

Performing a Taylor expansion of $G$ near
the $j$-th target yields 
\begin{equation}
\p_0 \sim G(\x_j,s|\x_0) + \nabla_{\x} G(\x,s|\x_0)\vert_{\x=\x_j} \cdot (\x-\x_j)+\frac{1}{2}{\bf H}_j\cdot(\x-\x_j)\otimes (\x-\x_j)+\ldots,
\end{equation}
where ${\bf H}_j $is the Hessian
\begin{equation}
H_j^{ab}=\left . \frac{\partial^2}{\partial x_a\partial x_b}G(\x,s|\x_0)\right \vert_{\x=\x_j} ,\quad a,b\in \{1,2,3\}.
\end{equation}
Introducing the stretched coordinates ${\bf y}=\epsilon^{-1}(\x-\x_j)$, we have
\begin{equation}
\label{p0}
\p_0 \sim G(\x_j,s|\x_0)+ \epsilon \nabla_{\x} G(\x_j,s|\x_0) \cdot \y +\frac{ \epsilon^2}{2} {\bf H}_j \cdot \y\otimes \y+\ldots
\end{equation}
 In stretched coordinates, equation (\ref{masterLT}) becomes
\begin{align}
	D\nabla_{\y}^2 P(\y,s) -s\epsilon^2 P(\y,s|\x_0)&= 0, \ |\y|>\ell_j;\
	P(\y,s)= 0,\ |\y|=\ell_j.
	\end{align}
Now consider a perturbation expansion of the inner solution around the $j$-th trap of the form $ p\sim P_0 + \epsilon P_1 + \epsilon^2P_2+O(\epsilon^3)$. This yields the hierarchy of equations (assuming $s\ll 1/\epsilon$)
\begin{subequations} 
\label{stretch}
\begin{align}
D\nabla_{\y}^2 P_n(\y,s) &=0;\ P_n(\y,s)= 0,\ |\y|=\ell_j,\ n=0,1,\\ 
	D\nabla_{\y}^2 P_n(\y,s) &=s\epsilon^2 P_{n-2}(\y,s)= 0 \ |\y|>\ell_j;
	P_n(\y,s)= 0,\ |\y|=\ell_j,\ \ n\geq 2.
	\end{align}
\end{subequations}
These are supplemented by far-field conditions obtained by matching with the near-field behavior of the outer solution.

Let us begin with the leading order contribution to the inner solution. Matching the far-field behavior of $P_0$ with the near-field behavior of $\p_0$ shows that
\begin{equation}
\Delta_y P_0 =0,\  |y|>1;\   P_0 \sim G(\x_j,s|\x_0) \mbox{ as } |y|\to \infty; \quad P_0=0 \mbox{ on } |y|=1. 
\end{equation}
This has the solution 
\begin{equation}
P_0 = G(\x_j,s|\x_0)( 1 - w(\y))),
\end{equation}
 with $w(\y)$ satisfying the boundary value problem
\begin{align}
\label{w}
\nabla_{\bf y}^2 w(\y)&=0,\  |\y|>\ell_j ; \quad w(\y)=1,\ |\y|=\ell_j;\ 
w(\y)\rightarrow 0\quad \mbox{as } |\y|\rightarrow \infty.\nonumber
\end{align}
In the case of a spherical target of radius $\ell_j$, we have
\begin{equation}
\label{w}
w(\y)=\frac{\ell_j}{|\y|}.
\end{equation}
It now follows that $\p_1$ satisfies equation (\ref{asym1}) together with the singularity condition
\[\p_1(\x,s)\sim - \frac{G_{j0}\ell_j}{|\x-\x_j|} \quad \mbox{as } \x\rightarrow \x_j,\]
where we have set $G_{j0}=G(\x_j,s|\x_0)$ and dropped the explicit dependence on $s,\x_0$ for notational convenience.
In other words, $\p_1(\x,s)$ satisfies the inhomogeneous equation
\begin{align}
 D\nabla^2 \p_1-s\p_1&={4\pi D} \sum_{j=1}^NG_{j0}\ell_j \delta(\x-\x_j),\, \x\in \calU;\quad
 \nabla \p_1\cdot \n=0,\ \x \in \partial \calU.
 \label{asym2}
\end{align}
This can be solved in terms of the modified Helmholtz Green's function:
\begin{equation}
\label{q1}
\p_1(\x,s)=- {4\pi}  D\sum_{j=1}^NG_{j0}\ell_jG(\x,s|\x_j).
\end{equation}

\subsection{Calculation of $P_1$} We now match the far-field behavior of $P_1$ with the $O(\epsilon)$ term in the expansion of $\p_0$, see equation (\ref{p0}), together with the near field behavior of $\p_1 $ around the $j$-th target. The latter takes the form
\begin{align*}
\p_1(\x,s)&\sim -\frac{G_{j0}\ell_j}{|\x-\x_j|}- {4\pi}D G_{j0} \ell_jR(\x_j,s|\x_j) -4\pi D\sum_{k\neq j}^NG_{k0}\ell_kG(\x_j,s|\x_k).
\end{align*}
It follows that
\begin{align}
P_1(\y,s)\rightarrow  \nabla_{\x} G(\x_j,s|\x_0) \cdot \y -4\pi D\sum_{k=1}^NG_{k0}\ell_k{\mathcal G}_{jk} \mbox{ as } |\y|\rightarrow \infty,
\end{align}
where
${\mathcal G}_{ij} =G(\x_i,s|\x_j)$ for $i\neq j$, and ${\mathcal G}_{ii} =R(\x_i,s|\x_i)$. Decompose the solution around the $j$-th target as $P_1=A_j^{(1)}+B_j^{(1)}$ with 
\[A_j^{(1)}\rightarrow \chi_j^{(1)}=-4\pi D\sum_{k=1}^NG_{k0} \ell_k{\mathcal G}_{jk} \mbox{ as } |\y|\rightarrow \infty,\]
and
\[B_j^{(1)}\rightarrow {\bf b}_j\cdot \y \mbox{ as } |\y|\rightarrow \infty,\ {\bf b}_j= \nabla_{\x} G(\x_j,s|\x_0) .\]
The solution for $A_j^{(1)}$ is then 
\begin{equation}
A_j^{(1)}(\y)=\chi_j^{(1)}(1-w(\y)),
\end{equation}
with $w(\y)$ given by equation (\ref{w}). We thus find 
\begin{equation}
A_j^{(1)}(\y)=\chi_j^{(1)} \left (1-\frac{\ell_j}{|\y|}\right ).
\end{equation}
In order to determine $B_j^{(1)}$, we introduce local spherical polar coordinates such that ${\bf b}_j=(0,0,b_j)$ and $\y\cdot {\bf b}_j=b_jr\cos \theta$, $0\leq \theta \leq \pi$. In spherical polar coordinates we have
\begin{align}
&\frac{\partial^2B_j^{(1)}}{\partial r^2}+\frac{2}{r}\frac{\partial B_j^{(1)}}{\partial r} +\frac{1}{r^2\sin \theta}\frac{\partial }{\partial \theta}\left (\sin \theta\frac{\partial B_j^{(1)}}{\partial \theta} \right )=0,\  r>1,\\
&   B_j^{(1)} \sim b_jr\cos \theta  \mbox{ as } r \to \infty;\quad B_j^{(1)}=0 \mbox{ on } r=\ell_j. 
\end{align}
Recall that Laplace's equation in spherical polar coordinates has the general solution
\begin{equation}
\label{gen}
B(r,\theta,\phi) = \sum_{l\geq 0}\sum_{m=-l}^l \left (a_{lm}r^l+\frac{b_{lm}}{r^{l+1}}\right )P_l^m(\cos \theta)\e^{im\phi},
\end{equation}
where $P_l^m(\cos \theta)$ is a Legendre polynomial. Imposing the Dirichlet boundary condition and the far-field condition implies that
\begin{equation}
\label{B1}
B_j^{(1)}(\y)=b_j\ell_j\cos \theta \left (\frac{|\y|}{\ell_j}-\frac{\ell_j^2}{|\y|^2}\right ).
\end{equation}
(This will contribute to the far-field behavior of the $O(\epsilon^3)$ term in the outer solution.)

\subsection{Calculation of $P_2$} Turning to the $O(\epsilon^2)$ terms, the outer contribution $\p_2$ satisfies equation (\ref{asym1}) supplemented by the singularity condition
\[\p_2(\x,s)\sim - \frac{\chi_j^{(1)}\ell_j}{|\x-\x_j|}, \quad \mbox{as } \x\rightarrow \x_j.\]
Using the same steps as in the derivation of $\p_1(\x,s)$, we obtain the result
\begin{equation}
 \p_2(\x,s)= -{4\pi}D\sum_{k=1}^N\chi_k^{(1)}\ell_kG(\x,s|\x_k) .
\end{equation}
Similarly, in order to calculate the inner contribution $P_2$, we have to match the far-field behavior of $P_2$ with the $O(\epsilon^2)$ term in the expansion of $\p_0$, see equation (\ref{p0}), the $O(\epsilon)$ terms in the expansion of $\p_1$,  together with the near field behavior of $\p_2 $ around the $j$-th target. That is,
\begin{align}
P_2(\y,s)\rightarrow  \frac{1}{2}{\bf H}_j\cdot \y\otimes \y-\nabla \tilde{p}_1 \cdot \y-4\pi D\sum_{k=1}^N\chi_k^{(1)}\ell_k{\mathcal G}_{jk} 
\end{align}
as $ |\y|\rightarrow \infty $. In addition, from equation (\ref{stretch}) we have
\begin{equation}
D\Delta_{\y} P_2 =sP_0,\  |y|>1;\quad P_2=0 \mbox{ on } |\y|=1. 
\end{equation}
 Again we decompose the inner term as
$P_2=A_j^{(2)}+B_j^{(2)}$ with 
\begin{align}
\label{Aj}
\Delta_yA_j^{(2)}&=0, |y|>1; \  A_j^{(2)}=0 ,\mbox{ on } |\y|=1, \\   A_j^{(2)}&\rightarrow \chi_j^{(2)}=-4\pi D\sum_{k=1}^N\chi_k^{(1)}\ell_k{\mathcal G}_{jk}\mbox{ as } |\y|\rightarrow \infty,\nonumber\end{align}
and
\begin{align}
\label{Bj}
D\Delta_{\y}B_j^{(2)}&=sP_0, |y|>1;\  B_j^{(2)}=0 ,\mbox{ on } |\y|=1,\\ B_j^{(2)}&\rightarrow \frac{1}{2}{\bf H}_j\cdot \y\otimes \y -\nabla \tilde{p}_1 \cdot \y\mbox{ as } |\y|\rightarrow \infty.\nonumber
\end{align}
The solution for $A_j^{(2)}$ is thus 
\begin{equation}
A_j^{(2)}(\y,s)=\chi_j^{(2)} \left (1-\frac{\ell_j}{|\y|}\right ).
\end{equation}
The calculation of $B_j^{(2)}$ is more involved.

The term $\nabla \p_1\cdot \y$ does not contribute to the flux so we will concentrate on on the dominant quadratic contribution to the far field condition.  In spherical polars the latter takes the form $r^2\sigma_2/2$ with
\begin{align*}
\sigma_2&= H_{xx} \sin^2\theta\cos^2\phi+H_{yy} \sin^2\theta\sin^2\phi+H_{zz} \cos^2\theta \\
&\quad +2[H_{xy}\sin^2 \theta\cos\phi \sin \phi+H_{xz} \sin \theta\cos \theta \cos \phi +H_{yz}\sin \theta\cos \theta \sin \phi]\\
&=H_{xx} \sin^2\theta(\cos 2\phi+1)/2+H_{yy} \sin^2\theta(1-\cos 2\phi)/2
+H_{zz}(\cos^2\theta-1/3) \\
&\quad +H_{zz}/3 +H_{xy}\sin^2 \theta\sin2\phi +2H_{xz} \sin \theta\cos \theta \cos \phi +2H_{yz}\sin \theta\cos \theta \sin \phi.
\end{align*}
All terms can be expressed in terms of $l=1,2$ spherical harmonics except
\begin{align*}
\frac{1}{2}(H_{xx}+H_{yy})\sin^2\theta +\frac{H_{zz}}{3}&=\frac{1}{2}(H_{xx}+H_{yy})(1-\cos^2\theta) +H_{zz}/3\\
&=\frac{1}{3}(H_{xx}+H_{yy}+H_{zz})+(H_{xx}+H_{yy})(1/3-\cos^2\theta).
\end{align*}
Thus the only term that yields a non-zero contribution to the flux integral ($l=0$ harmonic) is
\[\frac{r^2}{2}\overline{\sigma}_2=\frac{r^2}{6}\Delta_y G(\x_i,s|\x_0)=\frac{r^2s}{6D}G(\x_i,s|\x_0),\]
which is precisely the far-field contribution to the particular solution of the equation $D\Delta_yB_j^{(2)}=sG$. Hence,
\begin{equation}
\label{B2}
B_j^{(2)}=\overline{B}_j^{(2)}+\mbox{ higher-order harmonics},\quad \overline{B}_j^{(2)}=\frac{s}{D}\left (\frac{r^2}{6}-\frac{r\ell_j}{2}\right )G(\x_j,s|\x_0),
\end{equation}
and we
write the $O(\epsilon^2)$ contribution to the inner solution as
\begin{equation}
\label{P2}
P_2\sim \chi_j^{(2)}\left (1-\frac{\ell_j}{|\y|}\right ) +\frac{s}{D}\left (\frac{r^2}{6}-\frac{r\ell_j}{2}\right )G(\x_j,s|\x_0)+\mbox{ higher-order harmonics}.
\end{equation}
Finally, note that the same basic form of the inner solution holds at $O(\epsilon^n)$ with
\begin{equation}
\label{Bn}
P_n\sim \chi_j^{(n)}\left (1-\frac{\ell_j}{|\y|}\right ) + \overline{B}_j^{(n)} + \mbox{ higher-order harmonics}.
\end{equation}
Here $\overline{B}_j^{(n)}$ is the $(\theta,\phi)$ independent contribution generated by the particular solution of the equation
\begin{align*}
D\Delta_yB_j^{(n)}&=sP_{n-2}, |y|>1,\  B_j^{(n)}=0 ,\mbox{ on } |\y|=1,
\end{align*}
which also generates the correct far-field behavior. 
Moreover, the coefficients $\chi_j^{(n)}$ satisfy the iterative equation
\begin{equation}
\label{it}
\chi_j^{(n+1)}=-4\pi D\sum_{k=1}^N\chi_k^{(n)}\ell_k{\mathcal G}_{jk},\quad n\geq 1.
\end{equation}

\subsection{The flux}

Having obtained an $\epsilon$ expansion of the inner solution, we can use it to determine a corresponding expansion of the flux into the $j$-th target:
\begin{align}
\widetilde{J}_j(\x_0,s)&= D\epsilon^2 \int_{|\y|=\ell_j} \nabla_{\x} P \cdot \n \ d\y\sim 
D \epsilon \int_{|\y|=\ell_j}\left [ \nabla_{\y} P_0+\epsilon  \nabla_{\y} P_1+\ldots \right ] \cdot \n \ dS_{\y},
\end{align}
where now $\n$ denotes the normal out of the sphere $\calU_j$.
Introducing spherical polar coordinates $(r,\theta,\phi)$ relative to the center of the spherical target and setting 
\begin{equation}
P_n(\y)=\chi_j^{(n)}\left (1-\frac{\ell_j}{r}\right )+\overline{B}_j^{(n)}(r) +\mbox{higher harmonics},
\end{equation}
we can rewrite the asymptotic expansion of the flux as
\begin{align}
\label{JLTn}
\widetilde{J}_j(\x_0,s)&\sim \epsilon \widetilde{J}_j^{(0)}(\x_0,s)+\epsilon^2 \widetilde{J}_j^{(1)}(\x_0,s)+\epsilon^3 \widetilde{J}_j^{(3)}(\x_0,s)+\ldots,
\end{align}
with
\begin{align}
\widetilde{J}_j^{(n)}&= D  \ell_j^2 \int_0^{2\pi}\int_0^{\pi} \left . \frac{\partial}{\partial r}\right \vert_{r= \ell_j} \left [-\frac{  \ell_j\chi_j^{(n)}}{r}+\overline{B}_j^{(n)}(r)\right ]\sin\theta d\theta d\phi \nonumber\\
&=4\pi D\ell_j \chi_j^{(n)}+4\pi D\ell_j^2  \left . \frac{d \overline{B}_j^{(n)}}{dr}\right |_{r=\ell_j}.
\end{align}
In particular, using the explicit expressions for $\overline{B}_j^{(2)}$, $\chi_j^{(1)}$ and $\chi_j^{(2)}$, the flux through the $j$-th target to $O(\epsilon^3)$ is
\begin{align}
\label{JLT1}
\widetilde{J}_j(\x_0,s)
&\sim 4\pi \epsilon D\ell_j\left (G(\x_j,s|\x_0)-4\pi \epsilon D\sum_{k=1}^NG(\x_k,s|\x_0)\ell_k{\mathcal G}_{jk}(s)\right .\\
&\quad + \left .(4\pi \epsilon D)^2\sum_{k,l=1}^NG(\x_l,s|\x_0) \ell_k\ell_l{\mathcal G}_{kl}(s) {\mathcal G}_{jk}(s)\right )+\frac{4\pi \epsilon^3 s\ell_j^3}{6}G(\x_j,s|\x_0)\nonumber\\
&\quad+O(\epsilon^4).\nonumber
\end{align}
Note that the term $\overline{B}_j^{(1)} =0$. This expansion will be valid provided that $s\ll 1/\epsilon$. However, great care must be taken in taking the limit $s\rightarrow 0$, due to the fact that the modified Helmholtz Green's function diverges in this limit.

\section{Small-$s$ expansion}

We would like to use equation (\ref{JLT1}) to generate corresponding asymptotic expansions for the splitting probabilities and conditional FPT moments by performing a second expansion in the Laplace variable $s$.
However, $G$ is singular in the small-$s$ limit:
\begin{equation}
\label{GGG}
G(\x,s|\x')=\frac{1}{s|\calU|}+\overline{G}(\x,\x')+sF(\x,s|\x'),
\end{equation}
where $F$ is a non-singular function of $s$ and $\overline{G}(\x,\x')$ is the Neumann Green's function for the diffusion equation:
\begin{subequations}
\label{G0}
\begin{eqnarray}
D\nabla^2 \overline{G}(\x;\x')=\frac{1}{|\calU|} -\delta(\x-\x'),\, \x\in \calU;\ \partial_n\overline{G}=0,\, \x \in \partial \calU,\\
\overline{G}(\x,\x')=\frac{1}{4\pi D|\x-\x'|}+\overline{R}(\x,\x'),\ \int_{\calU}\overline{G}(\x,\x')d\x=0,
\end{eqnarray}
\end{subequations}
with $\overline{R}(\x,\x')$ again corresponding to the regular part of the Green's function. Since the component $A_j^{(n)}$ of the inner solution at order $n$ is proportional to $\chi_j^{(n)}$ and the latter is the product of $n+1$ Green's functions, it follows that $A_j^{(n)}$ will have singularities of $O(s^{-n-1})$ for all $n\geq 1$. On the other hand, the components $B_j^{(n)}$ are non-singular for $n\leq 2$, see equations  (\ref{B1}) and (\ref{B2}), and the singularity is $O(s^{2-n})$ for $n\geq 3$.
Substitution of equation (\ref{GGG}) into equation (\ref{JLT1}) implies that
\begin{align}
&\widetilde{J}_j \sim 4\pi \epsilon D\ell_j\bigg[\frac{1}{s|\calU|}+\overline{G}_{j0}+sF_{j0}\nonumber \\
&\quad -4\pi \epsilon D \sum_k\ell_k \left (\frac{1}{s|\calU|}+\overline{G}_{k0}+sF_{k0}\right )\left (\frac{1}{s|\calU|}+\overline{\mathcal G}_{jk} +sF_{jk}\right )\bigg ]+O(\epsilon^3) \nonumber \\
&\sim 4\pi \epsilon D\ell_j \overline{G}_{j0}+\frac{ 4\pi   D \epsilon}{s|\calU|}\ell_j \left [ 1-4\pi \epsilon D \sum_k\ell_k \overline{G}_{k0}-4\pi \epsilon D \sum_k\ell_k \overline{\mathcal G}_{jk}\right ]\nonumber \\
&\quad - \left (\frac{ 4\pi   D \epsilon}{s|\calU|}\right )^2\ell_j \bar{\ell}-(4\pi\epsilon D)^2\ell_j\sum_k\ell_k \overline{G}_{k0}\overline{\mathcal G}_{jk}\nonumber \\
&\quad -\frac{(4\pi\epsilon D)^2}{|\calU|}\ell_j\sum_k\ell_k (\overline{F}_{k0}+\overline{F}_{jk} )+
4\pi \epsilon sD\ell_j\overline{F}_{j0}+O(\epsilon^3,s\epsilon^2),
\label{JLT2}
\end{align}
where $\bar{\ell}=\sum_k \ell_k$, $\overline{G}_{k0}=\overline{G}(\x_k,\x_0)$, $\overline{\mathcal G}_{kj}=\overline{G}(\x_k,\x_j)$ for $k\neq j$, and $\overline{\mathcal G}_{jj}=\overline{R}(\x_j,\x_j)$. We have also set
\[\overline{F}_{jk}=\lim_{s\rightarrow 0}F(\x_j,s|\x_k).\]
The $\epsilon$-expansion in equation (\ref{JLT2}) indicates the potential problem we have in taking the limit $s\rightarrow 0$. This is due to the fact that terms involving factors of $\epsilon/s$ will become arbitrarily large and thus lead to a breakdown of the $\epsilon$ expansion. 

\subsection{Summing divergent terms}
It turns out that we can proceed by treating equation (\ref{JLT2}), including higher-order terms, as a triple expansion in $\epsilon$, $s$ and $\Lambda$, with
\begin{equation}
\Lambda = \frac{ 4\pi   D \epsilon \bar{\ell}}{s|\calU|}.
\end{equation}
This then converts a subset of terms in the Green's function expansion of $\epsilon^{n+1}A_j^{(n)}$ to $O(\epsilon^{r} \Lambda^{n+1-r})$ terms, $0\leq r\leq n$, see equation (\ref{Thetn}) below. (A similar observation holds for $\epsilon^{n+1}B_j^{(n)}$, $n\geq 3$, except that its lower-order singularity structure means that terms of $O(\Lambda)$ first appear at $O(\epsilon^3)$.)  As we now show, at each order of $\epsilon$, we obtain infinite power series in $\Lambda$ that can be summed to remove all singularities in the limit $s\rightarrow 0$.

In order to understand the origins of the infinite series, we use the iterative equation (\ref{it}). This shows that at $n$-th order, $n\geq 1$,  we have to deal with products of the form
\begin{align}
\label{pro}
&4\pi \epsilon^{n+1}D\ell_j\chi_j^{(n)}=4\pi \epsilon D\ell_j (-4\pi \epsilon D)^{n}  \sum_{k_1,\dots ,k_{n}}\left (\prod_{j=1}^{n}\ell_{k_j}\right ) \left (\frac{1}{s|\calU|}+\overline{G}_{k_10}+sF_{k_10}\right )\\
&\left (\frac{1}{s|\calU|}+\overline{{\mathcal G}}_{k_2k_1}+sF_{k_2k_1} \right )\dots \left (\frac{1}{s|\calU|}+\overline{{\mathcal G}}_{k_{n}k_{n-1}}+sF_{k_nk_{n-1}} \right )\left (\frac{1}{s|\calU|}+\overline{{\mathcal G}}_{jk_{n}} +sF_{jk_n}\right ) .\nonumber
\end{align}
It is convenient to introduce the following quantities:
\begin{subequations}
\label{alpha}
\begin{align}
\alpha_j^{(1)}&=\frac{1}{\bar{\ell}^{n-1}}\sum_{k_1,\dots ,k_{n}}\left (\prod_{j=1}^{n}\ell_{k_j}\right ) \bigg ( \overline{G}_{k_10}+\overline{{\mathcal G}}_{k_2k_1}+\ldots +\overline{{\mathcal G}}_{k_nk_{n-1}} +\overline{{\mathcal G}}_{jk_{n}}  \bigg)\nonumber \\
&= \sum_{k=1}^N \ell_k \bigg \{\overline{G}_{k0}+   \overline{{\mathcal G}}_{jk} \bigg \}  +\frac{(n-1)}{\bar{\ell}} \sum_{l,k=1}^N \ell_l\ell_k \overline{{\mathcal G}}_{lk}  ,
\end{align}
\begin{align}
\alpha_j^{(2)}&=\frac{1}{\bar{\ell}^{n-2}}\sum_{k_1,\dots ,k_{n}}\left (\prod_{j=1}^{n}\ell_{k_j}\right ) \frac{1}{2}\bigg [\overline{G}_{k_10}\bigg ( \overline{{\mathcal G}}_{k_2k_1}+\ldots +\overline{{\mathcal G}}_{jk_{n}}  \bigg) \nonumber \\
&\quad + \overline{{\mathcal G}}_{k_2k_1}\bigg ( \overline{G}_{k_10}+\ldots +\overline{{\mathcal G}}_{jk_{n}}  \bigg)+\ldots + \overline{{\mathcal G}}_{jk_{n}}\bigg ( \overline{G}_{k_10}+\ldots +\overline{{\mathcal G}}_{k_{n}k_{n-1}}  \bigg)\bigg ]\nonumber \\
&=  \sum_{k,l=1}^N \ell_k \ell_l\bigg \{ \overline{G}_{k0}\overline{{\mathcal G}}_{kl} 
+\overline{G}_{k0}\overline{{\mathcal G}}_{jl} +\overline{{\mathcal G}}_{jk}\overline{{\mathcal G}}_{kl}
\bigg \}\nonumber \\
&\quad +\frac{n-2}{\bar{\ell}}\sum_{k,l,m=1}^N \ell_k \ell_l\ell_m \bigg \{  \overline{G}_{k0}\overline{{\mathcal G}}_{lm}+\overline{{\mathcal G}}_{jk}\overline{{\mathcal G}}_{lm} +  \overline{{\mathcal G}}_{kl}\overline{{\mathcal G}}_{lm}\bigg \} \nonumber\\
&\quad +\frac{a(n)}{\bar{\ell}^2}\sum_{k,l,m,m'=1}^N \ell_k \ell_l\ell_m\ell_{m'} \overline{{\mathcal G}}_{kl}\overline{{\mathcal G}}_{mm'} ,\nonumber \\
&\equiv \alpha_j^{(2,2)}+\frac{n-2}{\bar{\ell}}\alpha_j^{(2,3)}+\frac{a(n)}{\bar{\ell}^2}\alpha^{(2,4)},
\end{align}
\begin{align}
\beta_j^{(1)}&=\frac{1}{\bar{\ell}^{n-1}}\sum_{k_1,\dots ,k_{n}}\left (\prod_{j=1}^{n}\ell_{k_j}\right ) \bigg ( \overline{F}_{k_10}+\overline{F}_{k_2k_1}+\ldots +\overline{F}_{k_nk_{n-1}} +\overline{F}_{jk_{n}}  \bigg)\nonumber \\
&=  \sum_{k=1}^N \ell_k \bigg \{ \overline{F}_{k0}+ \overline{F}_{jk}  \bigg \} +\frac{(n-1)}{\bar{\ell}} \sum_{l,k=1}^N \ell_l\ell_k \overline{F}_{lk} .
\end{align}
\end{subequations}
Note that the terms involving the factor $(n-2)$ only appear when $n>2$, and 
\[a(n)=\frac{(n+1)n}{2}-3(n-1),\ n\geq 4,\quad a(n)=0,\ n<4.
\]
We have also introduced the following quadratic functions of the Green's functions:
\begin{subequations}
\begin{align}
\alpha_j^{(2,2)}&=\sum_{k,l=1}^N \ell_k \ell_l\bigg \{ \overline{G}_{k0}\overline{{\mathcal G}}_{kl} 
+\overline{G}_{k0}\overline{{\mathcal G}}_{jl} +\overline{{\mathcal G}}_{jk}\overline{{\mathcal G}}_{kl}
\bigg \},\\
\alpha_j^{(2,3)}&=\sum_{k,l,m=1}^N \ell_k \ell_l \ell_m \bigg \{ \overline{G}_{k0}\overline{{\mathcal G}}_{lm} +\overline{{\mathcal G}}_{jk}\overline{{\mathcal G}}_{lm} + \overline{{\mathcal G}}_{kl}\overline{{\mathcal G}}_{lm} \bigg \},\\
\alpha^{(2,4)}&=\sum_{k,l,m,m'=1}^N  \ell_k \ell_l\ell_m\ell_{m'} \overline{{\mathcal G}}_{kl}\overline{{\mathcal G}}_{mm'} .
\end{align}
\end{subequations}
One can interpret the index $n$ of $\alpha_j^{(n,m)}$ as the order of the homogeneous polynomial in $\overline{G}$, whereas $m$ indicates the number of independent summations over target indices. Such notation generalizes to higher-orders in $\epsilon$.

Collecting all terms involving $(s|\calU|)^{-n-1}$, $(s|\calU|)^{-n}$ and $(s|\calU|)^{-n+1}$ in equation (\ref{pro}) then yields
for $n\geq 1$
\begin{align}
4\pi \epsilon^{n+1}D\ell_j\chi_j^{(n)}&=- \frac{\ell_j }{\bar{\ell}} \left ( -\Lambda \right )^{n+1}+\frac{4\pi \epsilon D\ell_j}{\bar{\ell}} (-\Lambda)^{n } \alpha_j^{(1)}
-\frac{(4\pi \epsilon D)^2\ell_j}{\bar{\ell}} \left ( -\Lambda \right )^{n-1} \alpha_j^{(2)}
\nonumber \\
&\quad -\frac{(4\pi \epsilon D)^2\ell_j}{|\calU|}  \left ( -\Lambda \right )^{n-1} 
\beta_j^{(1)}+O(\Lambda^{n-2}\epsilon^3)+O(\epsilon^n s)
\label{Thetn}
\end{align}
Explicitly including all $O(\epsilon \Lambda^r)$ terms of equation (\ref{Thetn}), $r\geq 1$, in the asymptotic expansion (\ref{ee}) of the flux, we have
\begin{align}
&\widetilde{J}_j(\x_0,s)
\sim 4\pi \epsilon D\ell_j \overline{G}(\x_j,\x_0) \\
&\quad +\frac{\ell_j\Lambda}{\bar{\ell}} \sum_{m\geq 0}(-\Lambda)^m \left [ 1-4\pi \epsilon D \sum_k\ell_k \overline{G}(\x_k,\x_0)-4\pi \epsilon D \sum_k\ell_k \overline{G}(\x_j,\x_k)\right ]\nonumber \\
&\quad +\frac{4\pi \epsilon \ell_j D\Lambda^2}{\overline{\ell}^2} \sum_{m\geq 0}(m+1)(-\Lambda)^m\sum_{i=1}^N\ell_i\overline{\mathcal G}_{ij} \ell_j \nonumber \\
&\quad -(4\pi\epsilon D)^2\ell_j\sum_k\ell_k \overline{G}_{k0}\overline{\mathcal G}_{jk}+\epsilon^2 \Theta_j^{(2)} +
4\pi \epsilon sD\ell_j\overline{F}_{j0}+O(\epsilon^3,s\epsilon^2). \nonumber 
\end{align}
Here $\Theta_j^{(2)}$ collects all $O(\Lambda^{n-1}\epsilon^2)$ terms in equation (\ref{Thetn}) for $n>1$.
If we now formally sum the infinite series using 
\begin{subequations}
\label{insum}
\begin{align}
 \Lambda\sum_{m\geq 0}(-\Lambda)^m&=\Lambda(1-\Lambda+\Lambda^2 \ldots )=\frac{\Lambda}{1+\Lambda},\\
\Lambda^2\sum_{m\geq 0}(m+1)(-\Lambda)^m&=\Lambda^2(1-2\Lambda+3\Lambda^2\ldots )=\Lambda^2 \frac{d}{d\Lambda}\frac{\Lambda}{1+\Lambda}=\frac{\Lambda^2}{(1+\Lambda)^2} ,
\end{align}
\end{subequations}
we have
\begin{align}
\label{ee}
&\widetilde{J}_j(\x_0,s)
\sim 4\pi \epsilon D\ell_j \overline{G}(\x_j,\x_0)  + \frac{\ell_j}{\bar{\ell}}\frac{\Lambda}{1+\Lambda}\bigg [ 1-4\pi \epsilon D \sum_k\ell_k \overline{G}(\x_k,\x_0)\\
&\quad -4\pi \epsilon D \sum_k\ell_k \overline{G}(\x_j,\x_k)\bigg ]+\frac{4\pi \epsilon \ell_j D}{\overline{\ell}^2}\frac{\Lambda^2}{(1+\Lambda)^2}\sum_{i=1}^N\ell_i\overline{\mathcal G}_{ij} \ell_j \nonumber \\
&\quad -(4\pi\epsilon D)^2\ell_j\sum_k\ell_k \overline{G}_{k0}\overline{\mathcal G}_{jk}+\epsilon^2 \Theta_j^{(2)} +
4\pi \epsilon sD\ell_j\overline{F}_{j0}+O(\epsilon^3,s\epsilon^2).\nonumber
\end{align}
We can now safely take the limit $s\rightarrow 0$ with $\Lambda \rightarrow \infty$.

Let us now consider the sum over the $O(\Lambda^{n-1}\epsilon^2)$ terms, $n>1$:
\begin{align*}
\Theta_j^{(2)}&=\frac{(4\pi  D)^2\ell_j}{\bar{\ell}}\bigg [  \alpha_j^{(2,2)} \Lambda\sum_{m\geq 0}(-\Lambda)^m  -\alpha_j^{(2,3)}\frac{ \Lambda^2}{\bar{\ell}}\sum_{m\geq 0}(m+1)(-\Lambda)^m \\
&\quad +\alpha^{(2,4)}\frac{1}{\bar{\ell}^2}\sum_{m\geq 4}a(m)\Lambda^{m-1}-
\frac{\bar{\ell}}{|\calU|}\sum_{m\geq 0}(-\Lambda)^m \sum_{k=1}^N \ell_k \bigg \{F_{k0}+   F_{jk} \bigg \}  \\
&\quad +\frac{\Lambda}{|\calU|} \sum_{m\geq 0}(m+1)(-\Lambda)^m \sum_{l,k=1}^N \ell_l\ell_k F_{lk} \bigg ]
\end{align*}
All of the infinite $\Lambda$-series can be summed along the lines of equation (\ref{insum}). In particular,\begin{align*}
\sum_{m\geq 4}a(m)\Lambda^{m-1}&=\sum_{m\geq 4}\left [\frac{m(m+1)}{2}-3(m-1)\right ]\Lambda^{m-1}\\
&=\frac{1}{2}\frac{d^2}{d\Lambda^2}\frac{\Lambda^5}{1+\Lambda}-3\Lambda \frac{d}{d\Lambda}\frac{\Lambda^3}{1+\Lambda}\\
&=\frac{10\Lambda^3}{1+\Lambda}-\frac{5\Lambda^4}{(1+\Lambda)^2}+\frac{\Lambda^5}{(1+\Lambda)^3}-\frac{9\Lambda^3}{1+\Lambda}+\frac{3\Lambda^4}{(1+\Lambda)^2}\\
&=\frac{\Lambda^3}{1+\Lambda}-\frac{2\Lambda^4}{(1+\Lambda)^2}+\frac{\Lambda^5}{(1+\Lambda)^3}=\frac{\Lambda^3}{(1+\Lambda)^3}.
\end{align*}
Hence,
\begin{align}
\Theta_j^{(2)}&=\frac{(4\pi  D)^2\ell_j}{\bar{\ell}}\bigg [   \frac{\Lambda}{1+\Lambda}  \alpha_j^{(2,2)}-\frac{ 1}{\bar{\ell}}\frac{\Lambda^2}{(1+\Lambda)^2}\alpha_j^{(2,3)}+\frac{1}{\bar{\ell}^2} \frac{\Lambda^3}{(1+\Lambda)^3}\alpha^{(2,4)}\nonumber \\
&\quad 
-\frac{\bar{\ell}}{|\calU|}\frac{1}{1+\Lambda} \sum_{k=1}^N \ell_k \bigg \{F_{k0}+   F_{jk} \bigg \}   +\frac{\Lambda}{|\calU|} \frac{\Lambda}{(1+\Lambda)^2} \sum_{l,k=1}^N \ell_l\ell_k F_{lk} \bigg ].
\label{ee2}
\end{align}

\subsection{Splitting probabilities and conditional MFPTs}

 An asymptotic expansion of the splitting probability $\pi_j(\x_0)$ defined in equation (\ref{split}) can now be obtained by taking the limit $s\rightarrow 0$ in equations (\ref{ee}) and (\ref{ee2}):
\begin{align}
 \pi_j(\x_0)&=\lim_{s\rightarrow 0}\widetilde{J}_j(\x_0,s)\sim \frac{\ell_j}{\overline{\ell}} +4\pi \epsilon D \ell_j\left [\overline{G}(\x_j,\x_0)-\frac{1}{\overline{\ell}}\sum_{k=1}^N\ell_k\overline{G}(\x_k,\x_0)\right ] \nonumber \\
 &\quad  +\epsilon \overline{\chi}_j -(4\pi\epsilon D)^2\ell_j\sum_k\ell_k \overline{G}_{k0}\overline{\mathcal G}_{jk}+ \epsilon^2\overline{\Theta}_j^{(2)}
  +O(\epsilon^3),
\label{split2}
\end{align}
where
\begin{equation}
\overline{\chi}_j=-\frac{4\pi \ell_j D}{\overline{\ell}}\left [\sum_{k=1}^N\overline{\mathcal G}_{jk}\ell_k-\frac{1}{\overline{\ell}}\sum_{i=1}^N\ell_i\overline{\mathcal G}_{ij} \ell_j\right ],
\end{equation}
and
\begin{align}
\label{Th2}
\overline{\Theta}_j^{(2)} &=\frac{(4\pi  D)^2\ell_j}{\bar{\ell}}\bigg [   \alpha_j^{(2,2)}-\frac{ 1}{\bar{\ell}}\alpha_j^{(2,3)}+\frac{1}{\bar{\ell}^2} \alpha^{(2,4)}\bigg ]\nonumber \\
&=\frac{(4\pi \epsilon D)^2\ell_j}{\bar{\ell}}\bigg [  \sum_{k,l=1}^N \ell_k \ell_l\bigg \{ \overline{G}_{k0}\overline{{\mathcal G}}_{kl} 
+\overline{G}_{k0}\overline{{\mathcal G}}_{jl} +\overline{{\mathcal G}}_{jk}\overline{{\mathcal G}}_{kl}
\bigg \}\nonumber \\
&\quad -\frac{1}{\bar{\ell}}\sum_{k,l,m=1}^N \ell_k \ell_l \ell_m \bigg \{ \overline{G}_{k0}\overline{{\mathcal G}}_{lm} +\overline{{\mathcal G}}_{jk}\overline{{\mathcal G}}_{lm} + \overline{{\mathcal G}}_{kl}\overline{{\mathcal G}}_{lm} \bigg \}\nonumber\\
&\quad +\frac{1}{\bar{\ell}^2} \sum_{k,l,m,m'=1}^N  \ell_k \ell_l\ell_m\ell_{m'} \overline{{\mathcal G}}_{kl}\overline{{\mathcal G}}_{mm'} \bigg ].
\end{align}
The $O(1)$ and $O(\epsilon)$ terms are identical to the expansion derived in \cite{Cheviakov11} by directly solving the boundary value problem for the splitting probabilities. The appearance of quadratic terms in the Green's function at $O(\epsilon^2)$ has also been shown by solving the BVP for the splitting probabilities \cite{Ward20}. Note that the $O(s)$ terms $F_{jk}$ in the $s$-expansion of the Green's function do not contribute to the splitting probabilities, at least to $O(\epsilon^2)$. Summing both sides of equation (\ref{split2}) with respect to $j$, one can check that the $O(\epsilon)$ and $O(\epsilon^2)$ terms each cancel, consistent with the normalization $\sum_k\pi_k=1$. 

Now consider the asymptotic expansion of the conditional MFPT:
\begin{equation}
\pi_k(\x_0)T_k(\x_0)=- \lim_{s\rightarrow 0} \left .\frac{d\widetilde{J}_k(\x_0,s)}{ds}\right |_{s=0}.
\end{equation}
For simplicity, we consider the first three orders in the expansion, namely, $O(1/\epsilon)$, $O(1)$ and $O(\epsilon)$ terms. For these contributions the only $s$-dependence is via the $\Lambda$-dependence in equations (\ref{ee}) and (\ref{ee2}) together with the term  $sF_{j0}$. Next,
setting
\[\Lambda=\frac{\overline{\Lambda}}{s},\quad \overline{\Lambda}=\frac{ 4\pi   D \epsilon \bar{\ell}}{|\calU|},\]
it follows that
\begin{equation}
\frac{d}{ds}\frac{\Lambda^n}{(1+\Lambda)^n}=\frac{d}{ds}\frac{\overline{\Lambda}^n}{(s+\overline{\Lambda})^n}=-\frac{\overline{n\Lambda^n}}{(s+\overline{\Lambda})^{n+1}}\rightarrow -\frac{n}{\overline{\Lambda}} \mbox{ as } \ {s\rightarrow 0}.
\end{equation}
Hence, differentiating equations (\ref{ee}) and (\ref{ee2}) with respect to $s$ and taking $s\rightarrow 0$ gives
\begin{align}
\pi_k(\x_0)T_k(\x_0)&=\frac{\ell_j}{\bar{\ell}}\frac{|\calU|}{ 4\pi   D \epsilon \bar{\ell}}\left [ 1-4\pi \epsilon D \sum_k\ell_k \overline{G}(\x_k,\x_0)-4\pi \epsilon D \sum_k\ell_k \overline{\mathcal G}_{jk}\right ] \\
&\quad +\frac{4\pi \ell_j D}{\overline{\ell}^2}\frac{2|\calU|}{ 4\pi   D  \bar{\ell}}\sum_{i=1}^N\ell_i\overline{\mathcal G}_{ij} \ell_j\nonumber\\
&\quad +\frac{(4\pi   D)^2\ell_j}{\bar{\ell}}\frac{|\calU|}{ 4\pi   D \bar{\ell}}\bigg [   \alpha_j^{(2,2)}-\frac{ 2}{\bar{\ell}}\alpha_j^{(2,3)}+\frac{3}{\bar{\ell}^2} \alpha^{(2,4)}\bigg ]+O(\epsilon^2)-4\pi D\ell_j\overline{F}_{j0}.\nonumber
\end{align}
The $O(1/\epsilon)$ and $O(1)$ terms agree with the corresponding expressions derived in \cite{Coombs15} by solving the boundary value problem for the conditional MFPTs. 

\section{Example for finite $s$}
\begin{figure}[b!]
\begin{center} 
\includegraphics[width=5cm]{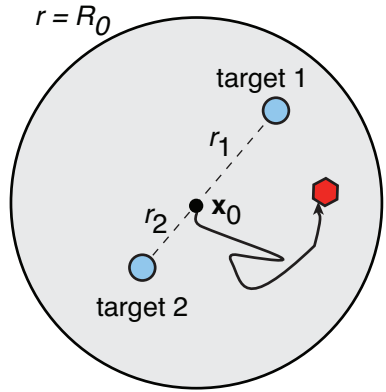} 
\caption{Spherical search domain of radius $r=R_0$ containing two diagonally opposed targets at distances $r_j$ from the center. The initial position of the searcher is taken to be at the center of the sphere, $\x_0=0$.}
\label{fig2}
\end{center}
\end{figure}

As an illustration of the above theory, suppose that the search domain is a sphere of radius $R_0$. Consider two targets of equal size $\ell_j=1$ located along a diagonal at a distance $r_j$ from the center, $j=1,2$, see Fig. \ref{fig2}. One of the advantages of a spherical domain is that the Green's function of the modified Helmholtz equation can be calculated explicitly \cite{Grebenkov20}:
\begin{align}
{G} (\x,s|\x_0)&=\frac{\e^{-\sqrt{s/D}|\x-\x_0|}}{4\pi |\x-\x_0|}-{G}_{\rm sp}(\x,s|\x_0)
\end{align}
with
\begin{align}
{G}_{\rm sp}(\x,s|\x_0)&=\frac{1}{4\pi}\sqrt{\frac{s}{D}}\sum_{n=0}^{\infty} (2n+1)P_n(\cos \theta)\frac{k_n'(\sqrt{s/D}R_0)}{i_n'(\sqrt{s/D}R_0)}\nonumber \\
&\qquad \times i_n(\sqrt{s/D}|\x|)
i_n(\sqrt{s/D}|\x_0|).
\label{Gsphere}
\end{align}
Here $P_n$ is a Legendre polynomial, $\x\cdot \x_0 =|\x||\x_0|\cos \theta$, and $i_n,k_n$ are modified spherical Bessel functions,
\begin{equation}
i_n(x)=\sqrt{\frac{\pi}{2x}}I_{n+1/2}(x),\quad k_n(x)=\sqrt{\frac{2}{\pi x}}K_{n+1/2}(x).
\end{equation}
In order to simplify the analysis, we will assume that the initial position $\x_0$ is at the center of the sphere. Using the identities
\begin{equation}
i_0(x)=\frac{\sinh x}{x},\quad k_0(x)=\frac{\e^{-x}}{x}, \quad i_n(0)=0, \ n>0,
\end{equation}
we see that
\begin{align}
G_{j0}&=\frac{e^{-\sqrt{s/D}|\x_j|}}{4\pi |\x_j|}-\frac{1}{4\pi}\sqrt{\frac{s}{D}}\frac{k_0'(\sqrt{s/D}R_0)}{i_0'(\sqrt{s/D}R_0)}i_0(\sqrt{s/D}|\x_j|).
\label{Gsphere2}
\end{align}
Similarly,
\begin{equation}
{\mathcal G}_{jk}=\frac{e^{-\sqrt{s/D}|\x_j-\x_k|}}{4\pi |\x_j-\x_k|}-{G}_{\rm sp}(\x_j,s|\x_k),\ j\neq k; \quad {\mathcal G}_{jj}=-\frac{1}{4\pi}\sqrt{\frac{s}{D}}-G_{\rm sp}(\x_j,s|\x_j).
\end{equation}
Further useful identities are
\begin{equation}
(2n+1) i'_n=ni_{n-1}+(n+1)i_{n+1},\quad -(2n+1) k'_n=nk_{n-1}+(n+1)k_{n+1}.
\end{equation}
The 3D Neumann Green's function for Laplace's equation in the sphere can also be written down explicitly so that \cite{Cheviakov11} 
\begin{align}
\overline{G}_{jk}=\overline{G}(\x_j,\x_k)&=\frac{1}{4\pi }\bigg [\frac{1}{|\x_j-\x_k|}+\frac{R_0}{|\x_j||\x_j'-\x_k|} +\frac{1}{2}(|\x_j|^2+|\x_k|^2) \nonumber \\
&\quad +\frac{1}{R_0}\ln \left (\frac{2R_0^2}{R_0^2-|\x_j||\x_k|\cos \theta+|\x_j||\x_j'-\x_k|}\right )\bigg ]-\frac{7}{10 \pi R_0},
\end{align}
for $j\neq k$, where $\x'=\x/|\x|^2$.
It also follows that
\begin{align}
\overline{G}_{jj}=\overline{R}(\x_j,\x_j)&=\frac{1}{4\pi }\bigg [\frac{R_0}{R_0^2-|\x_j|^2}+\frac{1}{R_0}\ln \left (\frac{R_0^2}{R_0^2-|\x_j|^2}\right )+|\x_j|^2 \bigg ]-\frac{7}{10 \pi R_0},
\end{align}
and
\begin{equation}
\overline{G}_{j0}=\frac{1}{4\pi}\left [\frac{1}{|\x_j|}+\frac{|\x_j|^2}{2}\right ] +1-\frac{7}{10 \pi R_0}.
\end{equation}

\begin{figure}[t!]
\begin{center} 
\includegraphics[width=8cm]{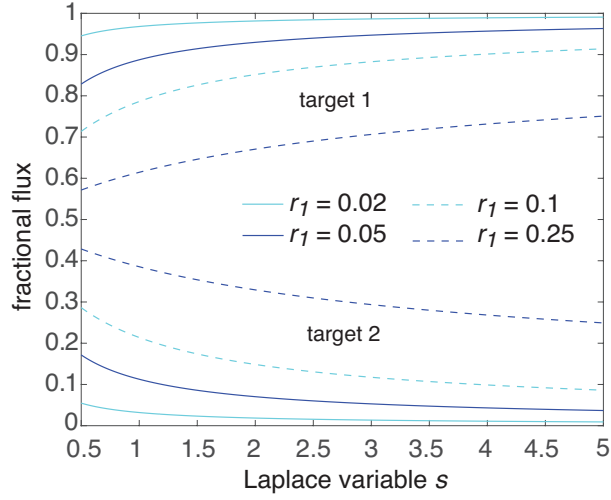} 
\caption{Plot of fractional target fluxes $\widetilde{j}_k(s)$, $k=1,2$, for the pair of targets shown in Fig. \ref{fig2}. The fluxes are determined by the asymptotic expansion given by equation (\ref{JLT1}). The distance of the second target is fixed at $r_2=0.5$ while the distance $r_1$ of the first target is varied. Other parameters values are $D=1$, $\ell_j=1$, $R_0=1$ and $\epsilon =0.01$.}
\label{fig3}
\end{center}
\end{figure}

We now have all the ingredients to determine the Laplace transformed fluxes $\widetilde{J}_k(s)$ according to equation (\ref{JLT1}) with $\x_0=0$. It is convenient to introduce the fractional fluxes
\begin{equation}
\label{fra}
\widetilde{j}_k(s)=\frac{\widetilde{J}_k(s)}{\widetilde{J}_1(s)+\widetilde{J}_2(s)},
\end{equation}
since the finite (truncated) asymptotic expansion does not blow up as $s \rightarrow 0$.
In Fig. \ref{fig3} we plot the fractional fluxes $\widetilde{j}_k(s)$, $k=1,2$, as a function of $s$ for $\epsilon \ll s \ll 1/\epsilon$. We also calculate the splitting probabilities $\pi_k$ using equation (\ref{split2}) and compare these values to $\lim_{s\rightarrow 0}\widetilde{j}_k(s)$ in Fig. 4. This illustrates the breakdown of the finite asymptotic expansion (\ref{JLT1}) in the limit $s\rightarrow 0$ due to singularities in the Green's function $G(\x,s|\y)$. As we showed in \S 4, one has to sum over an infinite number of terms in order to eliminate these singularities.

\begin{figure}[b!]
\begin{center} 
\includegraphics[width=8cm]{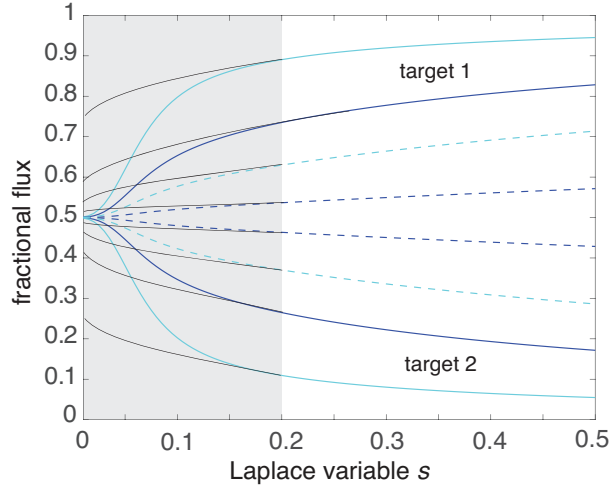} 
\caption{Same as Fig. \ref{fig3} except for the range of $s$. The curves added in the shaded region indicate an interpolation to the splitting probabilities $\pi_k$; the latter are calculated from equation (\ref{split2}). This illustrates the breakdown of the asymptotic expansion in the small $s$-regime due to $s$-singularities in the Green's function $G$.}
\label{fig4}
\end{center}
\end{figure}

In order to highlight one application of the above results, suppose that we include a stochastic resetting protocol into the search process given by equation (\ref{master}). That is, prior to being absorbed by one of the targets, the particle is allowed to instantaneously reset to a fixed location $\x_r$ at a random sequence of times generated by an exponential probability density $\psi(\tau)=r\e^{-r\tau}$, where $r$ is the resetting rate. The probability that no resetting has occurred up to time $\tau$ is then $\Psi(\tau)=1-\int_0^{\tau}\psi(s)ds=\e^{-r\tau}$. We identify $\x_r$ with the initial position by setting $\x_0=\x_r$. It turns out that the splitting probabilities and conditional FPT densities with resetting, which are distinguished by the subscript $r$, can be expressed in terms of the target fluxes without resetting as follows \cite{Bressloff20A}:
\begin{align}
\pi_{r,k}(\x_0)&=\frac{\pi_{k}(\x_0)-r\widetilde{\Pi}_{k}(\x_0,r)}{ 
1-r\widetilde{Q}(\x_0,r)}=\frac{\widetilde{J}_{k}(\x_0,r)}{\sum_{j=1}^N\widetilde{J}_{j}(\x_0,r)},
\label{Piee}
\end{align}
and
\begin{align}
\label{fcond}
\pi_{r,k}(\x_0)\widetilde{f}_{r,k}(\x_0,s)&=\frac{\pi_{k}(\x_0)-(r+s)\widetilde{\Pi}_{k}(\x_0,r+s)}{1-r\widetilde{Q}(\x_0,r+s)}=\frac{(r+s)\widetilde{J}_{k}(\x_0,r+s)}{s+r\sum_{j=1}^N\widetilde{J}_{j}(\x_0,r+s)}.
\end{align}
We have used equations (\ref{Pik}) and (\ref{QL}). Note, in particular, that the fractional fluxes $\widetilde{j}_k(s)$ with $s=r$, see equation (\ref{fra}), correspond precisely to the splitting probabilities with instantaneous resetting, $\pi_{r,k}$. Fig. \ref{fig3} then implies that as the resetting rate $r$ is increased, the probability of finding the target closest to the resetting point increases at the expense of the other target.

\section{Non-spherical target shapes}

So far the analysis has been restricted to spherically-shaped targets, since this avoids having to deal with multipole contributions to the inner solutions. However, as originally shown by Ward and Keller \cite{Ward93}, it is possible to generalize the asymptotic analysis of narrow capture problems to more general target shapes such as ellipsoids by applying classical results from electrostatics.
Here we briefly indicate how to apply such ideas to our asymptotic analysis of target fluxes, following a suggestion by Ward \cite{Ward20}. For simplicity, we only discuss contributions to the flux arising from the inner solution components $P_0$ and $P_1$. For a general target shape $\calU_j$, the solution to equation (\ref{stretch}a) for $n=0$ is of the form $P_0(\y,s)=G(\x_j,s|\x_0)(1-w_c(\y))$ for $\y\in \R^3\backslash \calU_j$,
with $w_c$ having the far-field behavior
\begin{equation}
w_c(\y)\sim \frac{C_j}{|\y|}+\frac{{\bf D}_j\cdot \y}{|\y|^3}+\ldots \mbox{as } |\y|\rightarrow \infty.
\end{equation}
Here $C_j$ is the capacitance and ${\bf D}_j$ the dipole vector of an equivalent charged conductor with the shape $\calU_j$. (For a sphere, $C_j$ is the radius $\ell_j$ and ${\bf D}_j=0$). Similarly, the solution to equation (\ref{stretch}a) for $n=1$ is
\begin{equation}
P_1(\y,s)={\bf b}_j\cdot {\bf W}_c(\y)+\chi_j^{(1)} (1-w_c(\y)),\ \y \in \R^3\backslash\calU_j,
\end{equation}
where ${\bf W}_c(\y)$ is a vector-valued function defined by
\begin{align}
\Delta_{\bf y}{\bf W}_c=0,\ \y \in \R^3\backslash\calU_j;\quad {\bf W}_c=0,\ \y \in \partial \calU_j;\ {\bf W}_c \sim \y+\frac{{\mathcal T}_j\y}{|\y|^3}+\ldots\mbox{ as } |\y|\rightarrow \infty.
\end{align}
Here ${\mathcal T}_j$ is the $3\times 3$ polarizability matrix for $\calU_j$. (It arises within classical electromagnetic theory when calculating the Rayleigh scattering of an object in the presence of an electric field with a specified gradient at infinity.)

It turns out that the contributions of $P_0$ and $P_1$ to the target flux can be obtained without explicitly solving for $w_c(\y)$ and ${\bf W}_c(\y)$. The basic idea is to apply the divergence theorem over the region $\R^3\backslash \calU_j$, which allows us to determine the flux in terms of the far-field behavior of each inner solution component. Define the domain $\calU_{\rho}=\{\y\, |\, |\y|\leq \rho\}$, where $\rho$ is sufficiently large so that $\calU_j \subset \calU_{\rho}$. The divergence theorem for a smooth function $F(\y)$ then implies
\begin{align}
\int_{\R^3\backslash \calU_j}\Delta_{\y}Fd\y=\lim_{\rho \rightarrow \infty} \int_{\calU_{\rho}\backslash \calU_j}\Delta_{\y}Fd\y=\lim_{\rho \rightarrow \infty} \int_{\partial \calU_{\rho}} \nabla_{\y}F\cdot \n dS_{\y}-\int_{\partial \calU_{j}} \nabla_{\y}F\cdot \n dS_{\y}.
\end{align}
We have taken the unit normal $\n$ to be pointing out of the domains $\calU_{\rho}$ and $\calU_j$. Applying this result to $P_0$ with $\Delta_{\y} P_0=0$, we have
\begin{align}
\widetilde{J}_0(\x_0,s)&=D\int_{|\y|=\ell_j} \nabla_{\y} P_0\cdot \n \ dS_{\y}=\lim_{\rho \rightarrow \infty} D\int_{\partial\calU_{\rho}}\nabla_{\y} P_0\cdot \n \ dS_{\y}\nonumber \\
&=-4\pi DC_jG_{j0} \lim_{\rho \rightarrow \infty} \rho^2\partial_{\rho}\rho^{-1}=4\pi DC_j G_{j0}.
\end{align}
Similarly,
\begin{align}
\widetilde{J}_1(\x_0,s)&=\lim_{\rho \rightarrow \infty} D\int_{\partial\calU_{\rho}}\nabla_{\y} P_1\cdot \n \ dS_{\y}\\
&=-4\pi DC_j\chi_{j}^{(1)} \lim_{\rho \rightarrow \infty} \rho^2\partial_{\rho}\rho^{-1}=4\pi DC_j \chi_j^{(1)}.\nonumber 
\end{align}
Hence, the $O(\epsilon)$ and $O(\epsilon^2)$ contributions to the flux in equation (\ref{JLT1}) carry over to more general target shapes under the substitution $\ell_j\rightarrow C_j$. High-order contributions can be handled in a similar fashion, although the analysis becomes more involved.

\section{Discussion} In this paper we used asymptotic PDE methods to calculate target fluxes for the 3D narrow capture problem. The latter concerns the diffusion of a particle in a bounded 3D domain with one or more small $O(\epsilon)$ absorbing interior targets or traps. Matching inner and outer solutions of the diffusion equation in Laplace space, we derived an explicit expression for the Laplace transformed flux $\widetilde{J}_j(s)$  into the $j$-th target, see equation (\ref{JLT1}), valid up to $O(\epsilon^3)$. The resulting asymptotic expansion involved sums over products of the Green's function $G(\x,s|\x')$ of the modified Helmholtz equation. One major motivation for focusing on the fluxes $\widetilde{J}_j(s)$ is that they act as generators of statistical quantities such as splitting probabilities and conditional FPT moments. However, this requires eliminating Green's function singularities that arise in the limit $s\rightarrow 0$. We showed how to achieve this by considering a triple expansion in $\epsilon$, $s$ and $\Lambda\sim \epsilon /s$. The singularities were then removed by performing partial summations over infinite power series in $\Lambda$. A second reason for considering the fluxes $\widetilde{J}_j(s)$ is that for finite $s$ they play an important role in extended narrow capture problems such as search processes with stochastic resetting, as we previously established for 2D search processes \cite{Bressloff20A,Bressloff20B}. In the example of \S 5, we briefly illustrated how to
 use the results of the current paper to explore analogous problems in 3D. 
 
\section*{Acknowledgement} PCB would like to express his thanks to Michael J. Ward (University of British Columbia) for his many insightful and helpful comments during the completion of this work.

\end{document}